\documentclass[a4paper,UKenglish,autoref]{lipics-v2021}
\hideLIPIcs  
\pdfoutput=1  
\nolinenumbers 
\makeatletter
\g@addto@macro\ps@headings{\let\@oddfoot\@empty}
\makeatother
\usepackage{booktabs}
\usepackage[frozencache]{minted}
\usepackage{tikz}
\usetikzlibrary{positioning,fit,arrows.meta,backgrounds}

\newcommand{\isabelle}{Isabelle/HOL}
\newcommand{\thingol}{Thingol}
\newcommand{\afp}{AFP}
\newcommand{\code}[1]{\texttt{#1}}

\definecolor{isabellebase}{HTML}{FFF2CC}
\definecolor{pythonbase}{HTML}{D5E8D4}
\colorlet{isabelleband}{isabellebase!50!white}
\colorlet{pythonband}{pythonbase!50!white}
\definecolor{isaboxbg}{HTML}{F7F4ED}
\setminted{style=default, breaklines=false, bgcolorpadding=4pt}
\newminted[pycode]{python}{fontsize=\footnotesize, xleftmargin=1em, bgcolor=pythonband}
\newminted[pycodesmall]{python}{fontsize=\footnotesize, bgcolor=pythonband}
\newminted[isacodesmall]{isabelle}{fontsize=\footnotesize, escapeinside=@@, bgcolor=isabelleband}
\newcommand{\codegap}{\vspace{-5pt}}
\newmintinline[isa]{isabelle}{bgcolor=none, breaklines=true}
\newmintinline[py]{python}{bgcolor=none, breaklines=true}
\newcommand{\isakw}[1]{\PYG{k}{#1}}
\DeclareUnicodeCharacter{03B1}{\ensuremath{\alpha}}
\DeclareUnicodeCharacter{03B2}{\ensuremath{\beta}}
\DeclareUnicodeCharacter{03BB}{\ensuremath{\lambda}}
\DeclareUnicodeCharacter{21D2}{\ensuremath{\Rightarrow}}
\DeclareUnicodeCharacter{2227}{\ensuremath{\wedge}}
\DeclareUnicodeCharacter{2228}{\ensuremath{\vee}}
\DeclareUnicodeCharacter{27F6}{\ensuremath{\longrightarrow}}
\DeclareUnicodeCharacter{2260}{\ensuremath{\neq}}
\DeclareUnicodeCharacter{00D7}{\ensuremath{\times}}
\DeclareUnicodeCharacter{21C0}{\ensuremath{\rightharpoonup}}

\addto\extrasUKenglish{%
}

\title{Python Code Generation for Isabelle/HOL}

\author{Nikolaus Huber}{University of Cambridge, UK}{nh661@cam.ac.uk}{https://orcid.org/0000-0002-1616-0602}{}
\author{Roly Perera}{University of Cambridge, UK}{}{https://orcid.org/0000-0001-9249-9862}{}
\author{Dominic Orchard}{University of Cambridge, UK \and University of Kent, UK}{}{https://orcid.org/0000-0002-7058-7842}{}

\authorrunning{N. Huber, R. Perera, and D. Orchard}

\Copyright{Nikolaus Huber, Roly Perera, and Dominic Orchard}



\supplementdetails[cite=huber_2026_24106993]{Software}{https://doi.org/10.5281/zenodo.24106993}

\funding{This research received financial support from Schmidt Sciences and Bloomberg.}

\acknowledgements{We thank the Institute of Computing for Climate Science for
providing a supportive working environment. Anthropic's Claude was utilised
during the design of the code generator backend, particularly for helping with
the integration into the wider Isabelle framework, as well as for planning out
and generating sections of the code.}

\EventEditors{John Q. Open and Joan R. Access}
\EventNoEds{2}
\EventLongTitle{42nd Conference on Very Important Topics (CVIT 2016)}
\EventShortTitle{CVIT 2016}
\EventAcronym{CVIT}
\EventYear{2016}
\EventDate{December 24--27, 2016}
\EventLocation{Little Whinging, United Kingdom}
\EventLogo{}
\SeriesVolume{42}
\ArticleNo{23}

\begin{document}

\maketitle

\begin{abstract}

The Isabelle theorem prover provides an environment for formalising mathematics
and reasoning about programs. To make verified implementations practically usable,
its code generator translates functions expressed in the executable fragment of
its logic into conventional programming languages. Its targets have traditionally
been functional languages (Standard ML, OCaml, Haskell, and Scala), recently
joined by Go. Python, which is becoming the dominant language of
scientific computing, data science, and education, has remained out of reach.
In this work, we close this gap with a Python backend for Isabelle's code generator.
This generator enables verified algorithms to be used
in Python applications, or to serve as verified references against which established
Python libraries can be differentially tested, without leaving the Python ecosystem.
We validate the backend on regression suites covering large parts of the HOL library
and by testing six verified data structures and algorithms against established
Python libraries. The generated code is slower than hand-written libraries, but well
suited as a verified reference.
\end{abstract}

\section{Introduction}\label{sec:intro}

Isabelle/HOL~\cite{nipkow2002} is an interactive theorem prover (ITP) based on
higher-order logic (HOL), a typed $\lambda$-calculus
with polymorphism and type classes. The user of Isabelle/HOL creates \emph{theory files},
which can include datatype and type class declarations, inductive definitions, and
functions. A fragment of HOL is \emph{executable}, which means that it can be
evaluated inside Isabelle, and also translated into programs in conventional
programming languages. Isabelle's code generation framework was reimplemented
by Florian Haftmann in his PhD thesis~\cite{haftmann2009,haftmann2010}, replacing
an earlier design~\cite{berghofer2002}, and has since featured a more
modular setup, in which parts of the generator can be shared by different
language backends. Traditionally, these backends covered functional programming
languages, such as Standard ML, OCaml, Haskell, and Scala, as their
evaluation schemes are also (at least loosely) based on the $\lambda$-calculus.
Recently, another backend~\cite{stubinger2025} for the programming language Go
has been added, which for the first time allowed generating code in an
imperative programming language.

Python is arguably one of the most widely used
programming languages right now. It has ranked among the top ten of the TIOBE
index for two decades, and has held first place since October 2021~\cite{tiobe2026}.
Its usage covers almost all areas of computing, and includes
scientific computing, data science, machine learning and AI, scripting,
automation, testing, and web backends. It is also one of the
most widely used programming languages in education, and often the first language
students learn. However, Python is still not a common target for code
generation from ITPs. To close this gap, we implemented a Python backend for
Isabelle's code generator. We hope that the possibility of dropping
verified implementations of core algorithms into established Python
applications will lower the barrier to adopting verified code. Moreover,
generated code can serve as a verified reference against which existing
Python libraries can be tested.

Several of Python's design choices simplify code generation from HOL. Its integers
have arbitrary precision, matching HOL's \isa{int}, and its \py{match} statement expresses
most of HOL's patterns directly. Since Python is dynamically typed, polymorphism
and type classes need not be fitted into a static type system.
Python, however, raises obstacles of its own:
CPython, the reference implementation of the Python language~\cite{python-glossary},
limits recursion depth and does not eliminate tail calls, closures capture
variables rather than values, and \py{if}/\py{else} and pattern matching are statements,
not expressions. \autoref{sec:translation} describes how the backend addresses
these issues.

In summary, this paper makes the following contributions:
\begin{itemize}
  \item A Python backend for Isabelle/HOL's code generator, available as an
    Isabelle component (\autoref{sec:translation}).
  \item A library setup that maps HOL types to native Python equivalents, 
    while preserving HOL's semantics
    where Python's built-in operations differ (\autoref{sec:library}).
  \item A validation of the backend using regression suites covering large parts
    of the HOL library, together with a performance evaluation on CPython and
    PyPy (\autoref{sec:evaluation}).
  \item A demonstration that generated code can serve as a verified reference
    for testing established Python libraries, through differential testing of
    six verified data structures and algorithms, including against SymPy's
    Hermite and Smith normal form implementations (\autoref{sec:evaluation}).
\end{itemize}

The code for the backend is publicly available~\cite{huber_2026_24106993} under an
open-source licence.
The rest of this paper is structured as follows. \autoref{sec:background}
introduces Isabelle's code generator and Python as a target language.
\autoref{sec:translation} presents the translation of HOL constructs to Python,
including the library setup for native types and the translation of tail calls
into loops. \autoref{sec:evaluation}
validates the backend and evaluates its performance. \autoref{sec:related}
surveys related work, \autoref{sec:future} discusses limitations and future work, and
\autoref{sec:conclusion} concludes.

\section{Background}\label{sec:background}

\subsection{The Isabelle Code Generator}

\begin{figure}[ht]
  \centering
  \scalebox{1.2}{\input{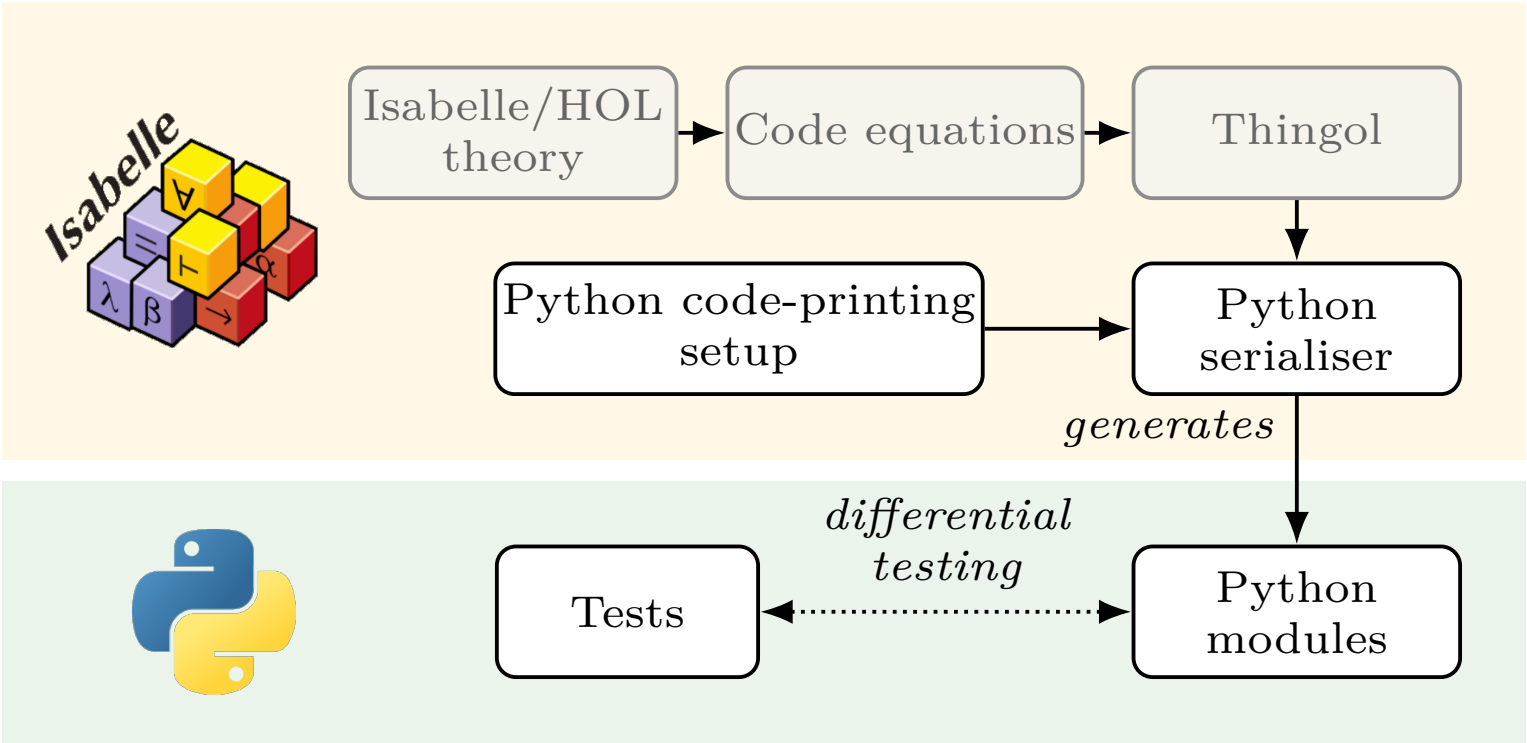}}
  \caption{The Python backend within the Isabelle code generator. Code equations are
    translated into a \thingol{} program, which the new serialiser prints as a Python
    module, guided by the \isa{code_printing} declarations of the library setup
    theories. The existing serialisers share the same front end (greyed out).}
  \label{fig:pipeline}
\end{figure}

\autoref{fig:pipeline} shows the overall pipeline of generating code from
an Isabelle/HOL theory. A user implements functions and datatypes through different Isabelle/HOL
definitions. These definitions give rise to \emph{code equations}, i.e.,
equations whose left-hand sides pattern-match on datatype constructors.
Code equations are typically derived from definitions in the executable fragment
of HOL, but even certain non-executable definitions, such as inductive predicates,
can be turned into code equations through additional
translations~\cite{berghofer2009}. After a preprocessing step that rewrites
the equations, they are translated into a small intermediate language called
\emph{\thingol{}}~\cite{haftmann2010}. \thingol{} is an ML-like functional language
extended with type classes. Its programs consist of datatypes, type classes and
their instances, and functions defined by pattern-matching equations over
constants, applications, abstractions, and case expressions. The translation from
Isabelle/HOL to \thingol{} is part of the Isabelle distribution and is shared by all
code generation backends.

Each language-specific backend is a \emph{serialiser}, which translates each
\thingol{} construct into a corresponding construct of the target language and
prints the result. By default, the code originating from each theory becomes a
separate Python module; alternatively, all code can be collected in a single
module.

Part of the generation is controlled by theory files that set up the
generation of Python code for specific HOL types and constants through
dedicated \isa{code_printing} declarations.
This is further explained in \autoref{sec:library}.

Since code equations are proven theorems, the generated code is correct with
respect to the HOL definitions in the sense of partial correctness, provided
that the serialiser and the \isa{code_printing} declarations
are correct; both are part of the trusted code base. A fully verified translation
is possible, as demonstrated by the CakeML code generator~\cite{hupel2018}, at the
cost of considerably more effort. We return to this in \autoref{sec:future}.

\subsection{Python as a Target}\label{sec:python-target}

At first glance, Python might seem an odd choice as the target for a code generator
from a functional intermediate language such as \thingol{}. However, several of its
design choices actually help during translation.

\textit{Dynamic typing.} Python does not rely on a type checker to verify type
correctness before execution. This greatly simplifies the translation of polymorphic
functions, as parametric polymorphism is essentially free. Type classes, however,
still require explicit dictionaries, and type errors in the generated code only surface
when it is executed.

\enlargethispage{\baselineskip}
\textit{Pattern matching and dataclasses.} While Python does not offer true algebraic
data types, it does include \emph{dataclasses}, which can be \emph{frozen} to make
instances immutable. Frozen dataclasses also automatically implement structural
equality and hashing. Since Python~3.10, the language additionally includes structural
pattern matching through \py{match}/\py{case} statements~\cite{pep634}, and
dataclass instances can be matched on in order to access their fields, 
as in \py{case Some(n)}. This allows us
to express most of \thingol{}'s pattern matching directly, without transforming it into
nested conditionals.

\textit{Integers and functions.} Python's integers have arbitrary precision, matching
HOL's integer semantics without additional library support, as is needed in other
backends such as OCaml. Even though Python is predominantly an imperative language,
it offers anonymous functions through \py{lambda} expressions and treats functions
as first-class values.

However, Python also has idiosyncrasies that have to be dealt with during translation.

\textit{Statements and expressions.} Unlike the functional target languages, Python
distinguishes between \emph{expressions} and \emph{statements}, and putting a statement
where an expression is expected (e.g., as the body of a \py{lambda}) leads to a
syntactically ill-formed program. In particular, \py{match}, \py{if}, and
assignments are statements, which \autoref{sec:translation} accounts for.

\textit{Recursion.} CPython does not eliminate tail calls. Together with its default
recursion limit of 1000 stack frames, this quickly leads to errors, as \thingol{} does not
include a loop construct and relies on recursion instead. \autoref{sec:tailcalls}
shows how we handle simple cases of tail calls.

\textit{Division.} Python raises an exception on division by zero, whereas HOL defines
$a \mathbin{\mathrm{div}} 0 = 0$ and $a \mathbin{\mathrm{mod}} 0 = a$. The library setup
in \autoref{sec:library} preserves HOL's semantics.

\begin{figure}[ht]
\begin{pycodesmall}
x = 1
f = lambda: x
x = 2
print(f())  # prints 2, not 1
\end{pycodesmall}
  \caption{Late binding in Python: the closure \py{f} looks up \py{x}
    when it is called, not when it is created.}
  \label{fig:late-binding}
\end{figure}

\textit{Late binding.} A design choice that can surprise even seasoned programmers is
Python's late binding of variables. Take, for example, the program in
\autoref{fig:late-binding}, where the variable \py{x} is bound late (after the definition of \py{f}), and therefore
the output of the fragment is \py{2} rather than \py{1}. Closures thus capture
variables, not values, which must be accounted for when turning recursion into loops
(\autoref{sec:tailcalls}).

\section{Translating \thingol{} to Python}\label{sec:translation}

The translation from \thingol{} to Python happens in two phases. First, we
translate the \thingol{} constructs into an internal representation of the
corresponding Python definitions; then we pretty-print those definitions,
taking care of Python's indentation-based syntax.

To illustrate how HOL constructs are translated, we use the small example
theory in \autoref{fig:running-thy}. It describes binary search trees with
\isa{datatype α tree = Leaf | Node (α tree) α (α tree)}: a tree is either a
\isa{Leaf} or a \isa{Node l x r} with left subtree \isa{l}, value \isa{x}, and
right subtree \isa{r}.

The function \isa{ins} is the standard insertion into a search tree. The type
annotation \isa{α::linorder} on its first argument restricts
\isa{α} to instances of the sort \isa{linorder}, i.e., to linearly
ordered types. The function \isa{root_or} returns a singleton list containing
the value at the root of a tree, or a default value if the tree is a
\isa{Leaf}. It is introduced with \isa{definition}, which defines a constant
by a single, non-recursive equation, whereas \isa{fun} allows recursive
definitions by pattern matching over several equations, as in \isa{ins}.
The higher-order function \isa{fold_tree} folds over a tree, and
\isa{from_list} builds a tree by repeatedly inserting the elements of a list.

Finally, the type class \isa{weight} declares a single operation
\isa{calc_weight :: α ⇒ nat}, which assigns a natural number to each value of an
instance type. The instance for \isa{nat} takes a number as its own weight.
The instance for trees is parametric: \isa{tree :: (weight) weight} states that
\isa{α tree} is an instance of \isa{weight} whenever \isa{α} is, and defines the
weight of a tree as the sum of the weights of its elements, computed with
\isa{fold_tree}. The constant \isa{example_weight} applies \isa{calc_weight} to a tree
whose single element is itself a tree of natural numbers, i.e., to a value of
type \isa{nat tree tree}. Evaluating it therefore requires both instances: the
instance for trees at the outer and at the inner level, and the instance for
\isa{nat} for the innermost elements.

We will now demonstrate the translation of the various parts of this example theory,
showing the generated Python code for each. Note that we have slightly simplified
the presented Python code for brevity and readability: we removed
redundant parentheses, renamed variables, merged nested conditionals into
\py{elif}, and omitted the fallback cases discussed in \autoref{sec:datatypes}.
Unless stated otherwise, the examples use the full setup described in
\autoref{sec:library}.

\begin{figure}[ht]
\begin{isacodesmall}
datatype α tree = Leaf | Node (α tree) α (α tree)

fun ins :: (α::linorder) ⇒ α tree ⇒ α tree where
  ins x Leaf = Node Leaf x Leaf
| ins x (Node l y r) =
    (if x < y then Node (ins x l) y r
     @\isakw{else}@ if y < x then Node l y (ins x r)
     @\isakw{else}@ Node l y r)

definition root_or :: α ⇒ α tree ⇒ α list where
  root_or d t = [case t @\isakw{of}@ Leaf ⇒ d
                           | Node _ x _ ⇒ x]

fun fold_tree :: (β ⇒ α ⇒ β ⇒ β) ⇒ β ⇒ α tree
               ⇒ β where
  fold_tree f e Leaf = e
| fold_tree f e (Node l x r) =
    f (fold_tree f e l) x (fold_tree f e r)

definition from_list :: (α::linorder) list
                       ⇒ α tree where
  from_list xs = foldr ins xs Leaf

class weight = fixes calc_weight :: α ⇒ nat

instantiation nat :: weight begin
definition calc_weight_nat :: nat ⇒ nat where
  calc_weight_nat n = n
instance .. end

instantiation tree :: (weight) weight begin
definition calc_weight_tree :: α tree ⇒ nat where
  calc_weight_tree =
    fold_tree (λl x r. l + calc_weight x + r) 0
instance .. end

definition example_weight :: nat where
  example_weight =
    calc_weight (Node Leaf
                 (Node Leaf (7::nat) Leaf) Leaf)
\end{isacodesmall}
  \caption{Running example: binary search trees in Isabelle/HOL (abridged, and
    without the quotation marks around terms and types).}
  \label{fig:running-thy}
\end{figure}

\subsection{Datatypes}\label{sec:datatypes}

The datatype \isa{α tree} is translated into one dataclass per constructor, as
shown in \autoref{fig:tr-datatype}. The nullary constructor \isa{Leaf} has no
fields, while \isa{Node} has three, one for each constructor argument. Each
field is annotated with \py{Any}: dataclasses require an annotation, but Python
does not check annotations at run time, so polymorphic datatypes need no special
treatment. The price is that static type checkers such as mypy~\cite{mypy} cannot check the
generated code, whose type correctness therefore rests on the code generator
alone. We revisit this design choice in \autoref{sec:future}.

\begin{figure}[ht]
\begin{pycodesmall}
@dataclass(frozen=True)
class Leaf:
    pass
@dataclass(frozen=True)
class Node:
    a: Any
    b: Any
    c: Any
\end{pycodesmall}
  \caption{The datatype \isa{α tree} from \autoref{fig:running-thy} translated to one frozen
    dataclass per constructor.}
  \label{fig:tr-datatype}
\end{figure}

Each dataclass is declared with \py{frozen=True}, which makes its instances
immutable and generates structural equality (\py{==}) and hashing. Immutability
matters because functional data structures are persistent: an update must leave
older versions intact. Since we neither create a common base class (e.g.,
\py{class tree}) nor use precise type annotations, the generated code does not
reflect that \isa{tree} is closed, i.e., that \isa{Leaf} and \isa{Node} are its
only constructors. Consequently, the backend ends every \py{match} whose last
case is not a catch-all with a fallback case that raises an error. For
readability, we omit this case in the figures of this section; we discuss it in
\autoref{sec:future}.

Constructor arguments are unnamed in Isabelle/HOL, so the generated fields get
systematic names (\py{a}, \py{b}, \py{c}). These names are irrelevant for
pattern matching, as dataclasses support positional class patterns, e.g.,
\py{case Node(l, x, r)}.

Nullary constructors are instantiated anew at every use, as in
\py{Node(Leaf(), x, Leaf())}. This is semantically harmless, because equality is
structural, and it allows uniform class patterns.

\subsection{Functions and Pattern Matching}

Functions, introduced through \isa{fun} or \isa{definition} (among others),
are translated depending on their inner structure. Those with a single
equation that does not pattern-match on its arguments are translated to a plain
\py{def} in Python, such as \isa{root_or} in \autoref{fig:tr-functions-nested}.
A \isa{definition} without any arguments and without sort constraints (i.e., a
constant), such as \isa{example_weight} in \autoref{fig:caching}, is
translated into a cached function without parameters (\autoref{sec:caching}).

\begin{figure}[ht]
\begin{pycodesmall}
def ins(d, x, t):
    match (x, t):
        case (x, Leaf()):
            return Node(Leaf(), x, Leaf())
        case (x, Node(l, y, r)):
            less = (d.order_linorder.preorder_order
                    .ord_preorder.less)
            if less(x, y):
                return Node(ins(d, x, l), y, r)
            elif less(y, x):
                return Node(l, y, ins(d, x, r))
            else:
                return Node(l, y, r)
\end{pycodesmall}
\caption{Translation of \isa{ins}, a function defined by multiple equations.}
\label{fig:tr-functions-multi-eq}
\end{figure}

Multi-equation functions compile to a \py{match} over the
tuple of input arguments with one \py{case} per equation. This can
be seen in the translation of \isa{ins} in \autoref{fig:tr-functions-multi-eq}.
In the generated code, we match over the input value \py{x} and the old tree
\py{t}. The first equation \isa{ins x Leaf} then turns into \py{case (x, Leaf())}.
This illustrates the uniform class patterns mentioned in the previous
subsection: \py{Leaf()} is written identically in patterns and expressions.
A shared singleton object would avoid allocating a new \py{Leaf()} at each use
and would still match the class pattern \py{Leaf()}; we return to this in
\autoref{sec:future}.

The additional argument \py{d} is the \emph{instance dictionary} for the
linear order. We use it to perform the comparison \py{less} when inserting
the value \py{x} into the tree. This uses the well-known
\emph{dictionary passing}~\cite{wadler1989,haftmann2007}
technique for compiling type classes, also used by the Standard ML,
OCaml, and Go backends, as none of these languages has native type classes
either. \thingol{} already makes explicit which dictionary is passed where, so the
backend only has to print these instance dictionaries. The comparison \py{less}
belongs to the superclass \isa{ord} of \isa{linorder}, which is reached
through a chain of superclass dictionaries (\autoref{sec:typeclasses}).

\begin{figure}[ht]
\begin{pycodesmall}
def root_or(d, t):
    match t:
        case Leaf():
            x = d
        case Node(_, y, _):
            x = y
    return [x]
\end{pycodesmall}
  \caption{Translation of \isa{root_or}: the \isa{case} expression nested
    inside the list is hoisted out of it.}
  \label{fig:tr-functions-nested}
\end{figure}

A more interesting example is the translation of \isa{root_or}, which has a
\isa{case} inside a list constructor. The \isa{case} expression
is turned into a \py{match} in Python, which itself is a statement and not an
expression, and therefore cannot be used inside a Python list constructor
directly. The backend converts the term to A-normal
form~\cite{flanagan1993}: it hoists the \py{match} before the list construction,
and assigns the result of each branch to a new temporary variable \py{x}, which
the \py{return} then places in the list.

To embed a statement in an expression, the Go backend takes a different
approach: it wraps the statement in an anonymous function that is called
immediately~\cite{stubinger2025}. Python could emulate this with a nested
\py{def} that is called in place, but hoisting avoids the additional function
definition and call.

Care is needed with the short-circuiting connectives \isa{∧}, \isa{∨}, and
\isa{⟶}, which we print as Python's equally short-circuiting \py{and} and
\py{or} (\autoref{fig:tr-short-circuit}).
Their right operand is only evaluated if the
left one does not already determine the result, e.g., \isa{hd xs} in
\isa{xs ≠ [] ∧ hd xs = 0} is never evaluated for the empty list.

\begin{figure}[ht]
\begin{isacodesmall}
definition conj_eq :: nat list ⇒ bool where
  conj_eq xs = (xs ≠ [] ∧ hd xs = 0)
\end{isacodesmall}
\codegap
\begin{pycodesmall}
def conj_eq(xs):
    return not null(xs) and hd(xs) == 0
\end{pycodesmall}
  \caption{A conjunction is printed as Python's short-circuiting \py{and}.}
  \label{fig:tr-short-circuit}
\end{figure}

However, hoisting a \isa{case} out of the right operand would break this, as the hoisted statement
runs unconditionally. If the right operand contains a \isa{case}, we therefore
translate the connective into nested \py{match}es (see \autoref{fig:tr-short-circuit-case}, 
where a trivial \py{match} on a Boolean has been simplified to an \py{if},
as linters would otherwise flag it).

\begin{figure}[ht]
\begin{isacodesmall}
definition conj_case :: nat list ⇒ bool where
  conj_case xs =
    (xs ≠ [] ∧
      (case hd xs @\isakw{of}@ 0 ⇒ True | Suc _ ⇒ False))
\end{isacodesmall}
\codegap
\begin{pycodesmall}
def conj_case(xs):
    if not null(xs):
        if hd(xs) == 0:
            x = True
        else:
            x = False
    else:
        x = False
    return x
\end{pycodesmall}
  \caption{A conjunction whose right operand contains a \isa{case} expression is
    turned into a conditional, so that \isa{hd xs} is only evaluated for
    non-empty lists.}
  \label{fig:tr-short-circuit-case}
\end{figure}

\subsection{Higher-Order Functions and Partial Application}

As functions are first-class values in Python, translating higher-order
code is fairly straightforward. However, Isabelle/HOL is curried, whereas
Python passes all arguments of a call at once.

Calls that supply all arguments of a known constant are printed as ordinary
calls with multiple arguments, such as the recursive call
\py{fold_tree(f, e, l)} in \autoref{fig:tr-higher-order-fold-tree}. Parameters
such as \py{f}, however, are applied to one argument at a time.

\begin{figure}[ht]
\begin{pycodesmall}
def fold_tree(f, e, t):
    match (f, e, t):
        case (f, e, Leaf()):
            return e
        case (f, e, Node(l, x, r)):
            return (f(fold_tree(f, e, l))(x)
                     (fold_tree(f, e, r)))
\end{pycodesmall}
  \caption{Translation of \isa{fold_tree}: the parameter \py{f} is applied
    to one argument at a time.}
  \label{fig:tr-higher-order-fold-tree}
\end{figure}

\enlargethispage{\baselineskip}
Conversely, a constant that receives fewer arguments than it expects is turned
into a curried form using nested \py{lambda}s, as in \isa{from_list}
(\autoref{fig:tr-higher-order-from-list}). Here, the first argument to
\py{foldr} is \py{ins}, partially applied to the instance dictionary \py{d},
which still expects the value \py{x} to insert and the old tree \py{t}.
\isa{foldr} itself comes from Isabelle/HOL's \isa{List} theory, which the code
generator includes automatically, as \isa{from_list} depends on it. Its
defining equations only take the operation and the list as arguments and return
a closure that expects the start value, so the generated \py{foldr} has two
parameters, and \py{Leaf()} is passed to its result.

\begin{figure}[ht]
\begin{pycodesmall}
def from_list(d, xs):
    return foldr(lambda x: lambda t: ins(d, x, t),
                 xs)(Leaf())
\end{pycodesmall}
  \caption{Translation of \isa{from_list}: the partially applied \py{ins}
    becomes a curried lambda.}
  \label{fig:tr-higher-order-from-list}
\end{figure}

\subsection{Type Classes}\label{sec:typeclasses}

We have already seen that sort constraints become dictionary arguments, such
as \py{d} in \py{ins}. We now turn to how these dictionaries are defined and
built.

Each type class becomes a frozen dataclass with one \py{Callable} field per
operation and one field per superclass, which holds the dictionary of that
superclass (\autoref{fig:tr-classes}). For \isa{weight}, the dataclass has a
single field \py{calc_weight}.

\begin{figure}[ht]
\begin{pycodesmall}
@dataclass(frozen=True)
class weight:
    calc_weight: Callable
\end{pycodesmall}
  \caption{The type class \isa{weight} from \autoref{fig:running-thy} as a
    dictionary type.}
  \label{fig:tr-classes}
\end{figure}

Each instance becomes a function that returns its dictionary. The parametric
instance for trees, \py{weight_tree}, takes the dictionary for the element type
as its argument \py{d} (\autoref{fig:tr-instance-tree}). In
\isa{example_weight}, the dictionary for \isa{nat tree} is thus
\py{weight_tree(weight_nat())}, which is passed to \py{calc_weight_tree}, the
implementation of \isa{calc_weight} for trees (\autoref{fig:caching}).
Instances that take no dictionary arguments, such as \py{weight_nat}, are
cached, as we describe in \autoref{sec:caching}.

\begin{figure}[ht]
\begin{pycodesmall}
# instance method for tree
def calc_weight_tree(d, t):
    return fold_tree(lambda l: lambda x: lambda r:
                    l + d.calc_weight(x) + r, 0, t)

# returns tree instance of class weight
def weight_tree(d):
    return weight(
       calc_weight =
         lambda a: calc_weight_tree(d, a))
\end{pycodesmall}
  \caption{The parametric instance of \isa{weight} for trees.}
  \label{fig:tr-instance-tree}
\end{figure}

Superclass projections become chains of attribute accesses. In \py{ins}
(\autoref{fig:tr-functions-multi-eq}), the comparison \py{less} belongs to the
class \isa{ord}, which \isa{linorder} extends via \isa{order} and
\isa{preorder}; it is therefore reached as
\mintinline[bgcolor=none, breaklines=true, breakafter=.]{python}{d.order_linorder.preorder_order.ord_preorder.less}.
Each step of the chain selects the dictionary of the next superclass, so deeper
class hierarchies only lead to longer chains and need no further support from
the backend.

Certain sort constraints (such as \isa{linorder} in \isa{ins}) are used only to select the
correct comparison function for the respective type. For native types like
\py{int}, we could use Python's comparison operators directly instead.
We comment on this in \autoref{sec:future}.

\subsection{Caching}\label{sec:caching}

A constant without arguments is not translated into a module-level variable,
but into a function without parameters, which is called wherever the constant
is used, as for \isa{example_weight} in \autoref{fig:caching}. Importing a
generated module therefore evaluates nothing.
Modules follow theories, and code equations may refer to constants of later
theories, so generated modules can import each other cyclically; reading a
value of a module that is still being imported would fail.

To still evaluate each value only once, the backend decorates these functions
with \py{@functools.cache}. The same applies to instances that take no
dictionary arguments, such as the instance of \isa{weight} for \isa{nat}, whose
dictionary is thus built only once. Caching can be switched off with the option
\isa{python_cache}; we discuss its effect on performance in
\autoref{sec:performance}.

\begin{figure}[ht]
\begin{pycodesmall}
@functools.cache
def weight_nat():
    return weight(
        calc_weight =
          lambda a: calc_weight_nat(a))

@functools.cache
def example_weight():
    return calc_weight_tree(
        weight_tree(weight_nat()),
        Node(Leaf(),
             Node(Leaf(), 7, Leaf()),
             Leaf()))
\end{pycodesmall}
  \caption{The instance of \isa{weight} for \isa{nat} and the constant
    \isa{example_weight}: both are evaluated only once.}
  \label{fig:caching}
\end{figure}

\subsection{Library Setup and Native Types}\label{sec:library}

Like the Go backend, our backend consists of two parts. The first part
takes care of the general translation of \thingol{} into Python, and is
written in Isabelle's own implementation language Standard ML. The second
part is an Isabelle/HOL theory, which loads this translation code and
introduces \isa{code_printing} declarations that map HOL types and constants
to their Python counterparts. We provide two such setup theories, which a
user of our backend can choose from. As \isa{code_printing} declarations
are part of the trusted code base, the basic setup requires less trust.

In the basic setup, we only cover the essential parts of HOL, roughly
Booleans, the \isa{unit} type, and strings; numbers are represented by HOL's
own datatypes, i.e., natural numbers in Peano encoding and integers in
binary. This already allows generating code from a large part of the HOL
library. All other types are translated like any other datatype; in
particular, lists become the dataclasses \py{Nil} and \py{Cons} rather than
Python lists (\autoref{fig:lib-basic-lists}).

\begin{figure}[ht]
\begin{isacodesmall}
fun snoc :: α list ⇒ α ⇒ α list where
  snoc [] y = [y]
| snoc (x # xs) y = x # snoc xs y

\end{isacodesmall}
\codegap
\begin{pycodesmall}
@dataclass(frozen=True)
class Nil:
    pass
@dataclass(frozen=True)
class Cons:
    a: Any
    b: Any

def snoc(xs, y):
    match (xs, y):
        case (Nil(), y):
            return Cons(y, Nil())
        case (Cons(x, xs), y):
            return Cons(x, snoc(xs, y))
\end{pycodesmall}
  \caption{Lists with the basic setup: they are translated like any other
    datatype, and list literals become nested constructor calls.}
  \label{fig:lib-basic-lists}
\end{figure}

The full setup extends this with additional code printings: integers and
natural numbers are mapped to Python's native integer type, and HOL lists
and tuples to their native Python counterparts. For lists, the function
\isa{snoc} from \autoref{fig:lib-basic-lists} then operates on Python lists
(\autoref{fig:lib-full-lists}). A third, optional theory maps immutable
arrays (\isa{iarray}) to Python tuples, which gives constant-time indexing.

\begin{figure}[ht]
\begin{pycodesmall}
def snoc(xs, y):
    match (xs, y):
        case ([], y):
            return [y]
        case ([x, *xs], y):
            return [x] + snoc(xs, y)
\end{pycodesmall}
  \caption{The function \isa{snoc} from \autoref{fig:lib-basic-lists} with the
    full setup: HOL lists are mapped to Python lists.}
  \label{fig:lib-full-lists}
\end{figure}

Python's arbitrary-precision integers already match HOL's integer semantics
closely. However, HOL is total, and as such \isa{x div 0} is defined as
\isa{0}, and \isa{x mod 0} as \isa{x}. To align the semantics, we include a
custom printer for these two operators (\autoref{fig:lib-divmod}).

\begin{figure}[ht]
\begin{isacodesmall}
code_printing
  constant (div) :: int ⇒ int ⇒ int ⇀ (Python)
    "(lambda a, b: a // b if b else 0)(_, _)"
| constant (mod) :: int ⇒ int ⇒ int ⇀ (Python)
    "(lambda a, b: a % b if b else a)(_, _)"
\end{isacodesmall}
  \caption{Custom code printing for \isa{div} and \isa{mod} on integers in the
    full setup.}
  \label{fig:lib-divmod}
\end{figure}

Similarly, subtraction on natural numbers is printed as \py{max(0, a - b)},
as HOL's subtraction on \isa{nat} is truncated at \isa{0}, whereas Python's
integers would become negative.

HOL's option type \isa{datatype α option = Some α | None} does not receive a
custom printer and is translated like any other datatype.
Unwrapping \isa{Some}, so that \isa{Some x} becomes \py{x} and \isa{None}
becomes Python's \py{None}, would not be injective, as \isa{Some None} and
\isa{None} would both map to \py{None}. The backend's documentation lists all
custom printings.

\subsection{Tail Calls as Loops}\label{sec:tailcalls}

Purely functional programming languages, like \thingol{}, have no loop
construct and rely on recursion instead. To avoid the stack growth of deep
recursion, many functional languages use \emph{tail-call elimination}: a call
that is the last operation of a function reuses the current stack frame, as
nothing is left to do in the current context once it returns.

CPython does not eliminate tail calls and has a default recursion limit of
only 1000 stack frames, so even simple functions that recurse over lists raise
a \py{RecursionError} on fairly modest inputs. Take, for example, the function \isa{sum_acc} in
\autoref{fig:tailcall-default}, which sums the elements of a list using an
accumulator. Its default translation recurses once per list element, and
therefore fails for lists of 1000 elements.

\begin{figure}[ht]
\begin{isacodesmall}
fun sum_acc :: int list ⇒ int ⇒ int where
  sum_acc [] acc = acc
| sum_acc (x # xs) acc = sum_acc xs (acc + x)
\end{isacodesmall}
\codegap
\begin{pycodesmall}
def sum_acc(xs, acc):
    match (xs, acc):
        case ([], acc):
            return acc
        case ([x, *xs], acc):
            return sum_acc(xs, acc + x)
\end{pycodesmall}
  \caption{A tail-recursive function and its default translation, which
    recurses once per list element.}
  \label{fig:tailcall-default}
\end{figure}

To support deep tail recursion, the backend can optionally turn self tail
calls into loops. The body of the function is then wrapped inside a
\py{while True:} loop. Whenever the function calls itself in tail position,
the arguments are reassigned instead, and \py{continue} jumps back to the
beginning of the loop (\autoref{fig:tailcall-loop}). The reassignment is
simultaneous, so that all new arguments are computed from the old ones. The
transformed function therefore no longer grows the stack, however long its
input is.

\begin{figure}[ht]
\begin{pycodesmall}
def sum_acc(xs, acc):
    while True:
        match (xs, acc):
            case ([], acc):
                return acc
            case ([x, *xs], acc):
                xs, acc = xs, acc + x
                continue
\end{pycodesmall}
  \caption{The function from \autoref{fig:tailcall-default} after the
    transformation: the self tail call becomes a loop.}
  \label{fig:tailcall-loop}
\end{figure}

The transformation only applies to direct self tail calls; recursive calls in
other positions, and mutual recursion, are left unchanged. Moreover, a function
whose recursive call creates a closure remains recursive: as Python closures
capture variables rather than values (\autoref{fig:late-binding}), a closure
created in one iteration would observe the reassignments of the next. For such functions,
users can raise the recursion limit with \py{sys.setrecursionlimit}
(see also \autoref{sec:future}).

\subsection{Name Resolution}

Python's reserved words (e.g., \py{def}, \py{case}, \py{while}) and the
built-in functions used by our library setup (e.g., \py{len}, \py{map}) are
reserved, so the code generator never uses them as identifiers. Top-level name
clashes are resolved with trailing underscores, following PEP~8~\cite{pep8}.

\section{Evaluation}\label{sec:evaluation}

For evaluation, we look at three different but somewhat interconnected questions:

\begin{enumerate}[({RQ}1)]
  \item Does the backend accept a broad range of real \isabelle{} theories and
    produce valid Python code from them?
  \item Does the generated code behave as expected?
  \item How does the performance of generated code compare to hand-written
    Python code, and does the specific Python runtime that is used matter?
\end{enumerate}

\begin{table}[b]
  \centering
  \caption{Differential test harnesses.}
  \label{tab:harnesses}
  \small
  \begin{tabular}{@{}lll@{}}
    \toprule
    Harness & Isabelle source & Library \\
    \midrule
    Splay tree    & \afp{} \code{Splay\_Tree}           & \code{SortedDict} \\
    RBT map       & HOL-DS \code{RBT\_Map}              & \code{SortedDict} \\
    Leftist heap  & HOL-DS \code{Leftist\_Heap}         & \code{heapq} \\
    Binomial heap & HOL-DS \code{Binomial\_Heap}        & \code{heapq} \\
    Hermite NF    & \afp{} \code{Hermite}               & SymPy \\
    Smith NF      & \afp{} \code{Smith\_Normal\_Form}   & SymPy \\
    \bottomrule
  \end{tabular}
\end{table}

\subsection{Regression Test Suites (RQ1)}

Isabelle already contains a test session to stress-test existing backends. We
adapted the test setup of the Go backend~\cite{stubinger2025}, which draws in a
significant part of the full HOL library; our version generates Python code for
both the basic and full setup described in \autoref{sec:library}. This results
in about 200 Python modules per setup, with roughly 50{,}000 lines of code in
total. We check each generated Python module for syntactic correctness by
compiling it with CPython (\py{py_compile}). For both setups, our backend
produces syntactically valid Python code. As the session contains no unit
tests, this does not cover execution behaviour.

\subsection{Differential Testing (RQ2)}

The regression suites only show that the generated code is syntactically
valid. To check its behaviour, we use two forms of differential testing.

First, we extract over 200 small Isabelle examples to Standard ML and Python,
run both, and compare the outputs. All tests pass, assuming the Standard ML
backend is correct.

Second, we test verified data structures and algorithms from the
AFP~\cite{afp,afp-splay,afp-hermite,afp-smith} and Isabelle/HOL's functional
data structures library~\cite{nipkow2025} against established Python libraries
(\autoref{tab:harnesses}). Each harness feeds the same random inputs, generated
with Hypothesis~\cite{maciver2019}, to both sides and compares the results; for
data structures, it runs 300 random sequences of up to 60 operations, checking
agreement after every step. No mismatch was found, which also
indicates that the libraries agree with the verified specifications on the
tested inputs.

Verified Isabelle components can thus serve as test oracles for Python code,
or be generated into Python and used directly, if performance permits
(\autoref{sec:performance}).

\subsection{Performance (RQ3)}\label{sec:performance}

\begin{figure}[ht]
  \centering
  \makebox[\textwidth][c]{\includegraphics[width=160mm]{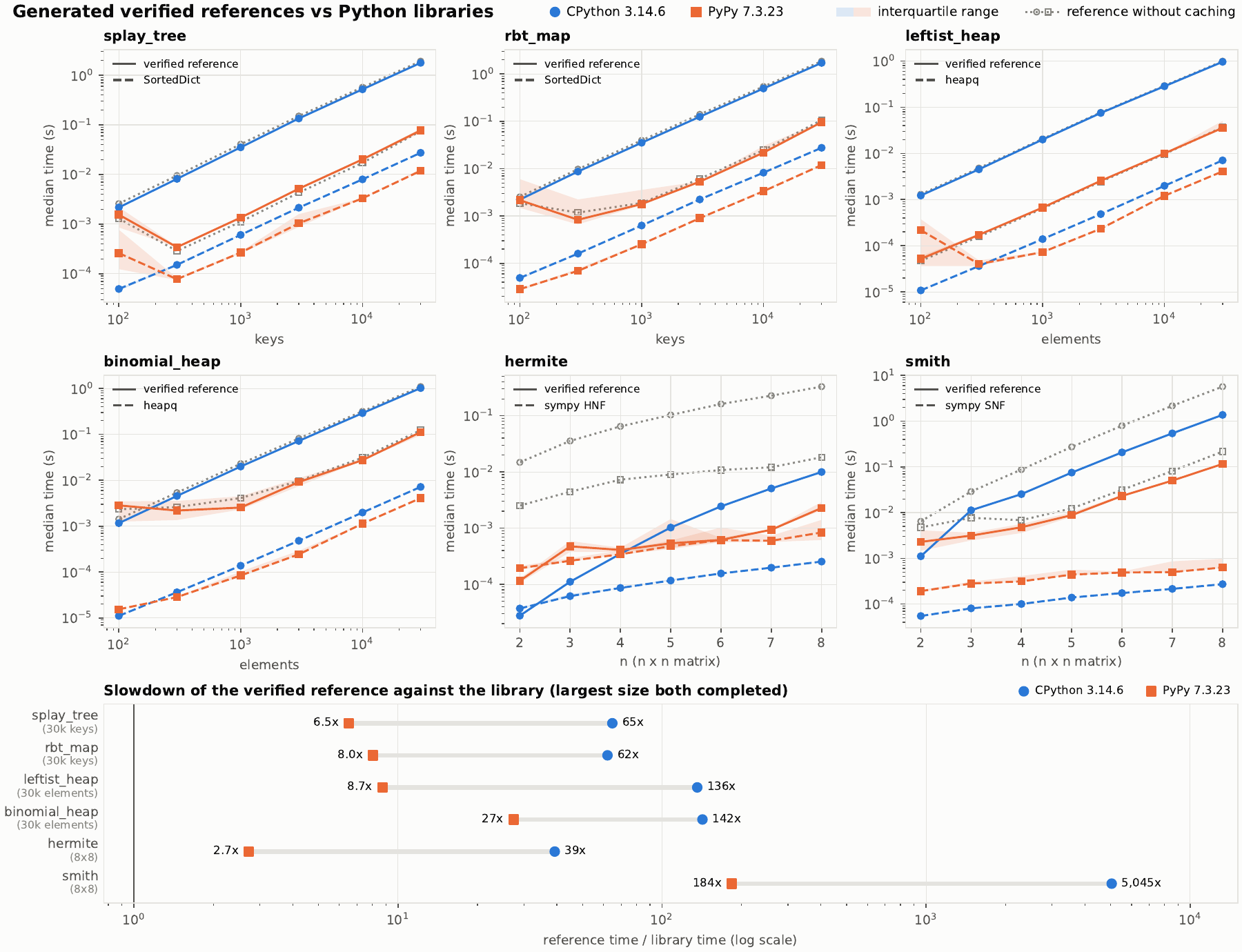}}
  \caption{Generated verified references vs.\ Python libraries on CPython~3.14.6 and
    PyPy~7.3.23. Top: median time against input size (log-log). Bottom: slowdown at the
    largest size both sides completed. Grey dotted: references built without caching.}
  \label{fig:bench}
\end{figure}

For the testing harnesses presented above, we investigated the performance
of each generated code module, and compared it to the respective established
Python library. We compared performance side by side on equivalent tasks, for 
example: insert $n$ keys into a map,
look all of them up, delete half of them, and list the remaining entries; insert 
$n$ elements into a heap, and then remove them one by one; apply normal form
algorithms to a random $n\times n$ integer matrix.
We compare two current implementations of Python,
the reference interpreter CPython (version 3.14.6), and PyPy (version 7.3.23).

CPython is arguably the most widely used implementation of the Python language, and
uses a bytecode interpreter under the hood. PyPy relies on JIT compilation to
translate frequently executed routines to machine code in order to speed up
execution along hot paths. All experiments were
run on a MacBook Pro M4 Max with 36\,GB of RAM.

For each input size, we report the median of 15 timed runs, preceded by 10
untimed warm-up runs to let PyPy's JIT compiler settle.
The interquartile range, i.e., the spread of the middle
half of the runs (25th--75th percentiles), is shown as shaded bands
around the median line in \autoref{fig:bench}.
We use the interquartile range rather than the standard deviation, as run times
are skewed by occasional slow runs, e.g., due to garbage collection or JIT compilation.

Unsurprisingly, the generated code is slower than the hand-written libraries (\autoref{fig:bench}).
For the four data structures, performance of generated code scales similarly to that of the equivalent
native Python libraries, but is 62--142$\times$ slower on CPython. JIT compilation seems
favourable for the code we are generating, as the performance gap is much
smaller when running PyPy, at 6.5--27$\times$. The Hermite normal form shows a
similar picture (39$\times$ and 2.7$\times$ at $8\times8$). The Smith normal
form is the outlier, with slowdowns of 5{,}045$\times$ on CPython and
184$\times$ on PyPy: the formalisation implements a generic algorithm for
Euclidean domains on top of a general matrix library, whose run time also
grows faster with the matrix size than SymPy's.

The grey dotted lines in \autoref{fig:bench} show the references built without
caching (\autoref{sec:caching}). For the data structures, caching makes little
difference: at the largest size, it speeds them up by 2--7\% on CPython, and on
PyPy, the difference (0.93--1.07$\times$) is within the measurement noise. The
matrix algorithms, which repeatedly use constants and dictionary-free instances,
profit considerably: at $5\times5$, caching speeds up the Hermite normal form by
100$\times$ on CPython and 18$\times$ on PyPy, and the Smith normal form by
3.6$\times$ and 1.3$\times$.

We believe that the exhibited slower performance is due to four reasons:

\textit{Operational overhead.} The generated code relies heavily on recursive function
calls, which are expensive in Python. Furthermore, pattern matching
is used throughout, which functional language runtimes or compilers often
optimise into fast decision trees~\cite{maranget2008a}, while Python's \py{match}
dynamically dispatches each case distinction. \autoref{tab:call-overhead}
shows the number of calls performed by the generated and the library code
for each harness on one example input. It clearly shows the call overhead
on all examples.

\textit{Immutability.} We rely on immutable data structures, and therefore
most of the operations we perform have to allocate new objects on the heap
on every update, while hand-written code can update objects on the heap
directly. This comes from the shift in paradigm, from functional \thingol{} to
imperative Python. The Objects column in \autoref{tab:call-overhead}
shows the number of dataclass instances created by the generated code. For
some tests (e.g., \code{splay\_tree}, \code{rbt\_map}) this amounts to over
40\% of the overall calls. This even includes constructors without arguments:
each comparison in the red-black tree, for instance, returns a fresh instance
of \py{LT}, \py{EQ}, or \py{GT}.

\begin{table}[t]
  \centering
  \caption{Function calls for one run of each workload, generated vs.\ library
    (CPython). Objects: dataclass instances created by the generated code.
  The `Ratio' column is the ratio of generated calls to library calls.}
  \label{tab:call-overhead}
  \small
\begin{tabular}{@{}lrrrrr@{}}
  \toprule
  & & \multicolumn{2}{c}{Generated} & Library & \\
  \cmidrule(lr){3-4}\cmidrule(lr){5-5}
  Benchmark & Size & Calls & Objects & Calls & Ratio \\
  \midrule
  splay\_tree & 1{,}000 & 305k & 129k & 12k & 25$\times$ \\
  rbt\_map & 1{,}000 & 223k & 104k & 12k & 19$\times$ \\
  leftist\_heap & 1{,}000 & 120k & 45k & 2.0k & 60$\times$ \\
  binomial\_heap & 1{,}000 & 129k & 23k & 2.0k & 65$\times$ \\
  hermite & $5\times5$ & 21k & 9 & 1.9k & 11$\times$ \\
  smith & $5\times5$ & 1.1M & 73k & 2.5k & 447$\times$ \\
  \bottomrule
\end{tabular}
\end{table}

\textit{Type classes.} Comparisons and arithmetic in the verified code are
generic over type classes. Instead of performing \py{x < y} on two integers,
the code therefore looks up the comparison in a dictionary argument and calls it,
sometimes through nested lambdas for partial application. In the red-black
tree, for instance, each comparison becomes
\mintinline[bgcolor=none, breaklines=true, breakafter=.]{python}{d[1].order_linorder.preorder_order.ord_preorder.less(x, y)}.
The libraries compare native values directly.

\textit{Native library code.} On CPython, \code{heapq} is implemented in C,
and \code{SortedDict} relies on Python's built-in lists, whose operations are
also implemented in C. This explains why the heaps show the largest slowdowns
on CPython. PyPy, in contrast, runs \code{heapq} as ordinary Python code, and
the gap for the leftist heap shrinks to 8.7$\times$.

Overall, this makes code generation with our backend more suitable for
differentially testing against established libraries than for using generated
code directly inside applications where performance is important.
While closing the performance gap completely will likely remain impossible,
we highlight some opportunities for optimisation in \autoref{sec:future}.

\subsection{Threats to Validity}

As already mentioned in \autoref{sec:library}, much of the code generation
pipeline is part of the trusted code base, and can only be validated through
proper testing. While the agreement in behaviour between the generated code for various
data structures and algorithms and established Python libraries gives a strong
indication of correctness, certainty can only be achieved if the whole pipeline
is verified down to the extracted Python code. We will remark on a possible
future extension towards a verified Python backend in \autoref{sec:future}.

Differential testing can only reveal disagreements on the inputs it
generates; it cannot prove their absence. Our selection of case studies
focuses on functional data structures and integer linear algebra, and does not
cover, e.g., floating-point or rational arithmetic. Finally, all performance
measurements were taken on a single machine, and parts of the compared
libraries are implemented in C, so the slowdowns reflect both the generated
code and the implementation language of the libraries.

\section{Related Work}\label{sec:related}

\textit{Isabelle backends.}
The standard Isabelle distribution ships with four code generation backends~\cite{haftmann2010},
targeting Standard ML, OCaml, Haskell, and Scala. The design is modular, so
additional backends can be added, reusing parts of the overall generator pipeline.
Recently, a new backend was added for the Go language~\cite{stubinger2025}, which
is now also part of the Archive of Formal Proofs~\cite{afp-go}. Go was the first
non-functional programming language added as a target, and thereby paved the
way for other imperative language backends, such as the one we present in this paper.
While we have taken inspiration from the Go backend, certain peculiarities of Python
also required different design choices. For example, while pattern matching translation
is easier in Python, as it features a dedicated \py{match} statement, statement
hoisting is different, as Python does not strictly follow static scoping rules.

\textit{Other provers and languages.} Other interactive theorem provers offer code generation from verified
formalisations as well. Rocq~\cite{therocqdevelopmentteam2025} can extract code to OCaml, Haskell, and
Scheme~\cite{letouzey2008}. Lean~4~\cite{moura2021} compiles to C. Nerodia, a fairly
new project, is a foreign function interface that allows writing native Python
modules in Lean~\cite{malone2026}. The programming language
Dafny~\cite{leino2010} already has a Python backend. Most of the code
expressed in Dafny is imperative and can be mapped directly to equivalent
Python. Pulse~\cite{ebner2025}, a proof-oriented language embedded in F*~\cite{swamy2011}, allows extraction of
programs to OCaml, C, and Rust.

Overall, very few provers or languages targeting
verified implementations currently cover Python.

\textit{Python verification.}
There are multiple tools which cover verification of Python code directly.
These range from type checkers such as mypy~\cite{mypy} to full static
verifiers like Nagini~\cite{eilers2018}. These are complementary approaches,
where Python code is verified post hoc, while our approach relies on
generating code from verified models.

\textit{Testing.}
Differential testing~\cite{mckeeman1998} compares several implementations of
the same specification on the same inputs, and property-based testing, as
introduced by QuickCheck~\cite{claessen2000a} and available for Python through
Hypothesis~\cite{maciver2019}, generates these inputs randomly. We combine both,
with the difference that one of the compared implementations is generated from
a verified formalisation, and thus serves as a reference for the established
library.

\section{Limitations and Future Work}\label{sec:future}

\textit{Recursion depth.}
Only direct tail calls of a function to itself are translated into loops
(\autoref{sec:tailcalls}). Mutual recursion and recursion that is not in tail
position still consume Python stack frames, and deep inputs, such as long
lists, run into CPython's recursion limit.  Trampolining~\cite{ganz1999},
as used by compilers to targets without tail-call elimination~\cite{tarditi1992,schinz2001},
possibly combined with a transformation into continuation-passing
style~\cite{flanagan1993}, would lift this restriction, at the cost of further
overhead per call.

\textit{Performance.}
\autoref{sec:performance} shows that the generated code is considerably slower
than hand-written libraries. Some of the overhead can be avoided within the
backend: constructors without arguments, such as \py{LT}, \py{EQ}, and \py{GT},
could be shared instances instead of being created anew for every use, and
type-class dictionaries for instances that are known at code generation
time could be specialised away, so that, e.g., a comparison on integers
becomes Python's native \py{<}. Native lists also come at a cost: patterns
such as \py{[x, *xs]} and concatenation (\py{[x] + xs}) copy the list, which
makes linear recursion over a list quadratic. A persistent list
representation, or offering both encodings, would avoid this. Finally, mapping
further HOL constants to Python built-ins in the full setup could reduce the
overhead, but also enlarge the trusted code base.

\textit{Type annotations.}
The generated code currently carries no type annotations beyond the fields
of dataclasses. Generating annotations according to PEP~484~\cite{pep484} would improve
integration with IDEs and allow static checking with tools such as
mypy~\cite{mypy}. Describing each datatype as a closed union of its
constructors would also make the fallback cases described in \autoref{sec:datatypes}
unnecessary, as a static type checker could then verify that the patterns are exhaustive. Extending our code generator to include
type annotations is left for future work.

\textit{Further types and code.}
The full setup maps integers and natural numbers to Python's \py{int}, but
does not cover \isa{rat} or \isa{real} yet; rationals could be mapped to
Python's \py{fractions.Fraction}. Monadic code in Imperative HOL, which
works on a mutable heap, could be mapped to mutable Python objects. 
This would avoid copying in data structures that are updated in place, 
and address some of the performance losses due to immutability described in 
\autoref{sec:performance}.

\textit{Verified backend.}
Like the other serialisers of Isabelle's code generator, our backend is part
of the trusted code base. Hupel and Nipkow showed for CakeML how this gap can
be closed with a verified compiler from Isabelle/HOL~\cite{hupel2018}. 
The code we generate largely targets PurePy~\cite{purepy2026},
a pure, functional subset of Python, which is designed to have a formal semantics.
Proving the translation into PurePy correct would make a verified pipeline
from Isabelle/HOL down to Python possible, 
following the approach of Hupel and Nipkow.

\section{Conclusion}\label{sec:conclusion}

In this paper, we have presented a Python backend for Isabelle/HOL's code generator.
It translates datatypes into immutable dataclasses, pattern-matching equations
into Python's structural pattern matching, and type classes into explicit
dictionaries. Its library setup maps HOL's numbers, lists, and immutable arrays
to native Python values, while preserving HOL's semantics where Python's
built-in operations differ, and self tail calls can optionally be translated
into loops.

Our evaluation shows that the backend produces valid Python code for a
large part of the HOL library. Differential testing of six verified data structures and
algorithms against established Python libraries revealed no disagreement,
further increasing trust in the correctness of our translation.
As expected, the generated code is slower than the hand-written libraries,
mainly due to call overhead, immutable data, and type-class dictionaries. The difference in performance
also depends on the particular Python runtime used, where PyPy seems
favourable to the style of code that is generated.

Even with slower performance, the generated code is suitable as a verified reference
implementation against which existing Python libraries can be tested. For use
inside applications, PyPy already narrows the gap, and the optimisations
outlined in \autoref{sec:future} could narrow it further. In the long run,
targeting a formally specified subset of Python such as PurePy could lead to a
verified pipeline from Isabelle/HOL down to Python code.

With increased calls for verification and formal methods in software
engineering, our approach provides an opportunity to leverage the
power of Isabelle in concert with one of the most popular languages,
Python, to increase trust in our software.

\bibliography{references,manual}

\begin{thebibliography}{10}

\bibitem{afp}
{Archive of Formal Proofs}.
\newblock \url{https://www.isa-afp.org}.
\newblock Accessed 2026-10-27.

\bibitem{berghofer2009}
Stefan Berghofer, Lukas Bulwahn, and Florian Haftmann.
\newblock Turning {{Inductive}} into {{Equational Specifications}}.
\newblock In Stefan Berghofer, Tobias Nipkow, Christian Urban, and Makarius
  Wenzel, editors, {\em Theorem {{Proving}} in {{Higher Order Logics}}}, volume
  5674 of {\em Lecture {{Notes}} in {{Computer Science}}}, pages 131--146.
  Springer Berlin Heidelberg, Berlin, Heidelberg, 2009.
\newblock \href {https://doi.org/10.1007/978-3-642-03359-9_11}
  {\path{doi:10.1007/978-3-642-03359-9_11}}.

\bibitem{berghofer2002}
Stefan Berghofer and Tobias Nipkow.
\newblock Executing {{Higher Order Logic}}.
\newblock In Paul Callaghan, Zhaohui Luo, James McKinna, Robert Pollack, Robert
  Pollack, Gerhard Goos, Juris Hartmanis, and Jan Van~Leeuwen, editors, {\em
  Types for {{Proofs}} and {{Programs}}}, volume 2277 of {\em Lecture {{Notes}}
  in {{Computer Science}}}, pages 24--40. Springer Berlin Heidelberg, Berlin,
  Heidelberg, 2002.
\newblock \href {https://doi.org/10.1007/3-540-45842-5_2}
  {\path{doi:10.1007/3-540-45842-5_2}}.

\bibitem{pep634}
Brandt Bucher and Guido van Rossum.
\newblock {PEP 634 -- Structural Pattern Matching: Specification}.
\newblock \url{https://peps.python.org/pep-0634/}, 2020.
\newblock {Python Enhancement Proposal}, accepted for {Python} 3.10.

\bibitem{claessen2000a}
Koen Claessen and John Hughes.
\newblock {{QuickCheck}}: A lightweight tool for random testing of {{Haskell}}
  programs.
\newblock In {\em Proceedings of the Fifth {{ACM SIGPLAN}} International
  Conference on {{Functional}} Programming}, pages 268--279. ACM, September
  2000.
\newblock \href {https://doi.org/10.1145/351240.351266}
  {\path{doi:10.1145/351240.351266}}.

\bibitem{afp-smith}
Jose Divas{\'o}n.
\newblock A verified algorithm for computing the {Smith} normal form of a
  matrix.
\newblock {\em Archive of Formal Proofs}, May 2020.
\newblock \url{https://isa-afp.org/entries/Smith_Normal_Form.html}.

\bibitem{afp-hermite}
Jose Divas{\'o}n and Jes{\'u}s Aransay.
\newblock {Hermite Normal Form}.
\newblock {\em Archive of Formal Proofs}, July 2015.
\newblock \url{https://isa-afp.org/entries/Hermite.html}.

\bibitem{ebner2025}
Gabriel Ebner, Guido Mart{\'i}nez, Aseem Rastogi, Thibault Dardinier, Megan
  Frisella, Tahina Ramananandro, and Nikhil Swamy.
\newblock {{PulseCore}}: {{An Impredicative Concurrent Separation Logic}} for
  {{Dependently Typed Programs}}.
\newblock {\em Proceedings of the ACM on Programming Languages},
  9(PLDI):1516--1539, June 2025.
\newblock \href {https://doi.org/10.1145/3729311} {\path{doi:10.1145/3729311}}.

\bibitem{eilers2018}
Marco Eilers and Peter M{\"u}ller.
\newblock Nagini: {{A Static Verifier}} for {{Python}}.
\newblock In Hana Chockler and Georg Weissenbacher, editors, {\em Computer
  {{Aided Verification}}}, volume 10981 of {\em Lecture {{Notes}} in {{Computer
  Science}}}, pages 596--603. Springer International Publishing, Cham, 2018.
\newblock \href {https://doi.org/10.1007/978-3-319-96145-3_33}
  {\path{doi:10.1007/978-3-319-96145-3_33}}.

\bibitem{flanagan1993}
Cormac Flanagan, Amr Sabry, Bruce~F. Duba, and Matthias Felleisen.
\newblock The essence of compiling with continuations.
\newblock {\em ACM SIGPLAN Notices}, 28(6):237--247, June 1993.
\newblock \href {https://doi.org/10.1145/173262.155113}
  {\path{doi:10.1145/173262.155113}}.

\bibitem{ganz1999}
Steven~E. Ganz, Daniel~P. Friedman, and Mitchell Wand.
\newblock Trampolined style.
\newblock In {\em Proceedings of the Fourth {{ACM SIGPLAN}} International
  Conference on {{Functional}} Programming}, pages 18--27, Paris France,
  September 1999. ACM.
\newblock \href {https://doi.org/10.1145/317636.317779}
  {\path{doi:10.1145/317636.317779}}.

\bibitem{haftmann2009}
Florian Haftmann.
\newblock {\em {Code Generation from Specifications in Higher-Order Logic}}.
\newblock PhD thesis, Technical University Munich, 2009.

\bibitem{haftmann2010}
Florian Haftmann and Tobias Nipkow.
\newblock Code {{Generation}} via {{Higher-Order Rewrite Systems}}.
\newblock In Matthias Blume, Naoki Kobayashi, Germ{\'a}n Vidal, David
  Hutchison, Takeo Kanade, Josef Kittler, Jon~M. Kleinberg, Friedemann Mattern,
  John~C. Mitchell, Moni Naor, Oscar Nierstrasz, C.~Pandu~Rangan, Bernhard
  Steffen, Madhu Sudan, Demetri Terzopoulos, Doug Tygar, Moshe~Y. Vardi, and
  Gerhard Weikum, editors, {\em Functional and {{Logic Programming}}}, volume
  6009 of {\em Lecture {{Notes}} in {{Computer Science}}}, pages 103--117.
  Springer Berlin Heidelberg, Berlin, Heidelberg, 2010.
\newblock \href {https://doi.org/10.1007/978-3-642-12251-4_9}
  {\path{doi:10.1007/978-3-642-12251-4_9}}.

\bibitem{haftmann2007}
Florian Haftmann and Makarius Wenzel.
\newblock Constructive {{Type Classes}} in {{Isabelle}}.
\newblock In Thorsten Altenkirch, Conor McBride, David Hutchison, Takeo Kanade,
  Josef Kittler, Jon~M. Kleinberg, Friedemann Mattern, John~C. Mitchell, Moni
  Naor, Oscar Nierstrasz, C.~Pandu~Rangan, Bernhard Steffen, Madhu Sudan,
  Demetri Terzopoulos, Doug Tygar, Moshe~Y. Vardi, and Gerhard Weikum, editors,
  {\em Types for {{Proofs}} and {{Programs}}}, volume 4502 of {\em Lecture
  {{Notes}} in {{Computer Science}}}, pages 160--174. Springer Berlin
  Heidelberg, Berlin, Heidelberg, 2007.
\newblock \href {https://doi.org/10.1007/978-3-540-74464-1_11}
  {\path{doi:10.1007/978-3-540-74464-1_11}}.

\bibitem{huber_2026_24106993}
Nikolaus Huber, Roly Perera, and Dominic Orchard.
\newblock {Python} code generation for {Isabelle/HOL} -- software artefact,
  November 2026.
\newblock \href {https://doi.org/10.5281/zenodo.24106993}
  {\path{doi:10.5281/zenodo.24106993}}.

\bibitem{hupel2018}
Lars Hupel and Tobias Nipkow.
\newblock A {{Verified Compiler}} from {{Isabelle}}/{{HOL}} to {{CakeML}}.
\newblock In Amal Ahmed, editor, {\em Programming {{Languages}} and
  {{Systems}}}, volume 10801 of {\em Lecture {{Notes}} in {{Computer
  Science}}}, pages 999--1026. Springer International Publishing, Cham, 2018.
\newblock \href {https://doi.org/10.1007/978-3-319-89884-1_35}
  {\path{doi:10.1007/978-3-319-89884-1_35}}.

\bibitem{mypy}
Jukka Lehtosalo et~al.
\newblock {mypy}: Optional static typing for {Python}.
\newblock \url{https://mypy-lang.org}.
\newblock {Accessed}: Oct. 26, 2026.

\bibitem{leino2010}
K.~Rustan~M. Leino.
\newblock Dafny: {{An Automatic Program Verifier}} for {{Functional
  Correctness}}.
\newblock In Edmund~M. Clarke and Andrei Voronkov, editors, {\em Logic for
  {{Programming}}, {{Artificial Intelligence}}, and {{Reasoning}}}, volume 6355
  of {\em Lecture {{Notes}} in {{Computer Science}}}, pages 348--370. Springer
  Berlin Heidelberg, Berlin, Heidelberg, 2010.
\newblock \href {https://doi.org/10.1007/978-3-642-17511-4_20}
  {\path{doi:10.1007/978-3-642-17511-4_20}}.

\bibitem{letouzey2008}
Pierre Letouzey.
\newblock Extraction in {{Coq}}: {{An Overview}}.
\newblock In Arnold Beckmann, Costas Dimitracopoulos, and Benedikt L{\"o}we,
  editors, {\em Logic and {{Theory}} of {{Algorithms}}}, volume 5028 of {\em
  Lecture {{Notes}} in {{Computer Science}}}, pages 359--369. Springer Berlin
  Heidelberg, Berlin, Heidelberg, 2008.
\newblock \href {https://doi.org/10.1007/978-3-540-69407-6_39}
  {\path{doi:10.1007/978-3-540-69407-6_39}}.

\bibitem{maciver2019}
David MacIver, Zac {Hatfield-Dodds}, and Many Contributors.
\newblock Hypothesis: {{A}} new approach to property-based testing.
\newblock {\em Journal of Open Source Software}, 4(43):1891, November 2019.
\newblock \href {https://doi.org/10.21105/joss.01891}
  {\path{doi:10.21105/joss.01891}}.

\bibitem{malone2026}
Mac Malone.
\newblock {Nerodia}.
\newblock \url{https://github.com/leanprover/nerodia}, 2026.
\newblock {Accessed}: Oct. 27, 2026.

\bibitem{maranget2008a}
Luc Maranget.
\newblock Compiling pattern matching to good decision trees.
\newblock In {\em Proceedings of the 2008 {{ACM SIGPLAN}} Workshop on {{ML}}},
  pages 35--46, Victoria BC Canada, September 2008. ACM.
\newblock \href {https://doi.org/10.1145/1411304.1411311}
  {\path{doi:10.1145/1411304.1411311}}.

\bibitem{mckeeman1998}
William~M. McKeeman.
\newblock Differential testing for software.
\newblock {\em Digital Technical Journal}, 10(1):100--107, 1998.

\bibitem{moura2021}
Leonardo~De Moura and Sebastian Ullrich.
\newblock The {{Lean}} 4 {{Theorem Prover}} and {{Programming Language}}.
\newblock In Andr{\'e} Platzer and Geoff Sutcliffe, editors, {\em Automated
  {{Deduction}} -- {{CADE}} 28}, volume 12699 of {\em Lecture {{Notes}} in
  {{Computer Science}}}, pages 625--635. Springer International Publishing,
  Cham, 2021.
\newblock \href {https://doi.org/10.1007/978-3-030-79876-5_37}
  {\path{doi:10.1007/978-3-030-79876-5_37}}.

\bibitem{afp-splay}
Tobias Nipkow.
\newblock {Splay Tree}.
\newblock {\em Archive of Formal Proofs}, August 2014.
\newblock \url{https://isa-afp.org/entries/Splay_Tree.html}.

\bibitem{nipkow2025}
Tobias Nipkow, editor.
\newblock {\em Functional {{Data Structures}} and {{Algorithms}}: {{A Proof
  Assistant Approach}}}.
\newblock ACM, New York, NY, USA, 1 edition, September 2025.
\newblock \href {https://doi.org/10.1145/3731369} {\path{doi:10.1145/3731369}}.

\bibitem{nipkow2002}
Tobias Nipkow, Markus Wenzel, Lawrence~C. Paulson, Gerhard Goos, Juris
  Hartmanis, and Jan Van~Leeuwen, editors.
\newblock {\em Isabelle/{{HOL}}}, volume 2283 of {\em Lecture {{Notes}} in
  {{Computer Science}}}.
\newblock Springer Berlin Heidelberg, Berlin, Heidelberg, 2002.
\newblock \href {https://doi.org/10.1007/3-540-45949-9}
  {\path{doi:10.1007/3-540-45949-9}}.

\bibitem{purepy2026}
{PurePy contributors}.
\newblock {PurePy: a functional subset of Python}, 2026.
\newblock {Accessed}: Oct. 27, 2026.
\newblock URL: \url{https://github.com/pure-py/pure-py-spec}.

\bibitem{python-glossary}
{Python Software Foundation}.
\newblock Python glossary: {CPython}.
\newblock \url{https://docs.python.org/3/glossary.html\#term-CPython}.
\newblock Accessed 2026-11-10.

\bibitem{schinz2001}
Michel Schinz and Martin Odersky.
\newblock Tail call elimination on the {{Java Virtual Machine}}.
\newblock {\em Electronic Notes in Theoretical Computer Science},
  59(1):158--171, November 2001.
\newblock \href {https://doi.org/10.1016/S1571-0661(05)80459-1}
  {\path{doi:10.1016/S1571-0661(05)80459-1}}.

\bibitem{afp-go}
Terru St{\"u}binger and Lars Hupel.
\newblock {Go} code generation for {Isabelle}.
\newblock {\em Archive of Formal Proofs}, January 2024.
\newblock \url{https://isa-afp.org/entries/Go.html}.

\bibitem{stubinger2025}
Terru St{\"u}binger and Lars Hupel.
\newblock Extending {{Isabelle}}/{{HOL}}'s {{Code Generator}} with {{Support}}
  for the {{Go Programming Language}}.
\newblock In Andre Platzer, Kristin~Yvonne Rozier, Matteo Pradella, and Matteo
  Rossi, editors, {\em Formal {{Methods}}}, volume 14934 of {\em Lecture
  {{Notes}} in {{Computer Science}}}, pages 3--19. Springer Nature Switzerland,
  Cham, 2025.
\newblock \href {https://doi.org/10.1007/978-3-031-71177-0_1}
  {\path{doi:10.1007/978-3-031-71177-0_1}}.

\bibitem{swamy2011}
Nikhil Swamy, Juan Chen, C{\'e}dric Fournet, Pierre-Yves Strub, Karthikeyan
  Bhargavan, and Jean Yang.
\newblock Secure distributed programming with value-dependent types.
\newblock In {\em Proceedings of the 16th {{ACM SIGPLAN}} International
  Conference on {{Functional}} Programming}, pages 266--278, Tokyo Japan,
  September 2011. ACM.
\newblock \href {https://doi.org/10.1145/2034773.2034811}
  {\path{doi:10.1145/2034773.2034811}}.

\bibitem{tarditi1992}
David Tarditi, Peter Lee, and Anurag Acharya.
\newblock No assembly required: Compiling standard {{ML}} to {{C}}.
\newblock {\em ACM Letters on Programming Languages and Systems},
  1(2):161--177, June 1992.
\newblock \href {https://doi.org/10.1145/151333.151343}
  {\path{doi:10.1145/151333.151343}}.

\bibitem{therocqdevelopmentteam2025}
{The Rocq Development Team}.
\newblock The {{Rocq Prover}}.
\newblock Zenodo, September 2025.
\newblock \href {https://doi.org/10.5281/ZENODO.15149628}
  {\path{doi:10.5281/ZENODO.15149628}}.

\bibitem{tiobe2026}
{TIOBE Software}.
\newblock {TIOBE Index} for {October} 2026.
\newblock \url{https://www.tiobe.com/tiobe-index/}, 2026.
\newblock {Accessed}: Oct. 25, 2026.

\bibitem{pep484}
Guido van Rossum, Jukka Lehtosalo, and {\L}ukasz Langa.
\newblock {PEP 484 -- Type Hints}.
\newblock \url{https://peps.python.org/pep-0484/}, 2014.
\newblock {Python Enhancement Proposal}, accepted for {Python} 3.5.

\bibitem{pep8}
Guido van Rossum, Barry Warsaw, and Alyssa Coghlan.
\newblock {PEP 8 -- Style Guide for Python Code}.
\newblock \url{https://peps.python.org/pep-0008/}, 2001.
\newblock {Python Enhancement Proposal}.

\bibitem{wadler1989}
P.~Wadler and S.~Blott.
\newblock How to make ad-hoc polymorphism less ad hoc.
\newblock In {\em Proceedings of the 16th {{ACM SIGPLAN-SIGACT}} Symposium on
  {{Principles}} of Programming Languages - {{POPL}} '89}, pages 60--76,
  Austin, Texas, United States, 1989. ACM Press.
\newblock \href {https://doi.org/10.1145/75277.75283}
  {\path{doi:10.1145/75277.75283}}.

\end{thebibliography}

\end{document}